\documentclass[aps,prd,twocolumn,noshowpacs,
nofootinbib,
superscriptaddress,groupedaddress, floatfix,superscriptaddress, preprintnumbers]{revtex4-2}
 \usepackage[utf8]{inputenc}
\usepackage{amsmath}
\usepackage{amsfonts}
\usepackage{graphicx}
\usepackage{amssymb}%
\usepackage{array,multirow}
\usepackage[breaklinks,colorlinks=true]{hyperref}
\begin{document}

\preprint{INT-PUB-22-023}

\title{Doubly heavy tetraquarks in the chiral quark soliton model}
%% Alphabetical order:
\author{Micha\l~Prasza\l owicz}
\email{michal.praszalowicz@uj.edu.pl}
\affiliation{Institute of Theoretical Physics, Jagiellonian University, S. \L ojasiewicza 11, 30-348 Krak\'ow, Poland}
\affiliation{Institute for Nuclear Theory, University of Washington, Box 351550, Seattle, Washington, 98195, USA}

\date{\today}

\begin{abstract}
 The chiral quark soliton model has been successfully applied to describe the heavy baryon spectrum, 
 both for charm and bottom, leading to the conclusion that the heavy quark has no effect on the soliton. This suggests that replacing a heavy
 quark by a heavy antidiquark $\bar{Q}\bar{Q}$ in color triplet should give a viable description of heavy tetraquarks. 
 We follow this strategy to compute tetraquark masses. To estimate heavy diquark masses we use the Cornell potential
 with appropriately rescaled parameters. The lightest charm tetraquark
 is 70~MeV above the $DD^*$ threshold. On the contrary, both nonstrange and strange bottom tetraquarks are bound
 by approximately 140 and 60~MeV, respectively. 
 \end{abstract}

\maketitle

\section{Introduction}

Recent discovery of a doubly charmed tetraquark ${\cal T}_{cc}^+$ 
with a mass of $\sim 3875$~MeV, approximately $\sim 300$~keV below the $D^{*+}D^0$ threshold,
by the 
LHCb Collaboration~\cite{LHCb:2021vvq,LHCb:2021auc} triggered a number
of theoretical studies of exotic heavy-light states. A comprehensive review of
multiquark states,
both experimental and theoretical, before ${\cal T}_{cc}^+$ discovery can be found in Ref.~\cite{Karliner:2017qhf}
and more recently after the discovery of ${\cal T}_{cc}^+$  in Ref.~\cite{Chen:2022asf} and references therein.
The up to date compilation of theoretical results is best illustrated in Fig.~42 of Ref.~\cite{Chen:2022asf}.

The existence of heavy tetraquarks has been anti{\-}cipated theoretically already many yers ago~\cite{Carlson:1987hh}.
In 1993, Manohar and Wise \cite{Manohar:1992nd} showed 
using heavy quark symmetry~\cite{Isgur:1991wq}
 that $QQq_1 q_2$
tetraquarks are bound in the limit $m_Q \rightarrow \infty$ (see also \cite{Cohen:2006jg,Cai:2019orb}).
This has been also pointed out more recently in Ref.~\cite{Eichten:2017ffp}.
To the best of our knowledge, the first estimate of a doubly heavy 
tetraquark mass is from Lipkin in 1986 \cite{Lipkin:1986dw}
(although the fourfold heavy tetraquarks were discussed even earlier in 1982 \cite{Ader:1981db}). 
We have reviewed the variational approach of Ref.~ \cite{Lipkin:1986dw} in Ref.~\cite{Praszalowicz:2019lje} 
adding new information coming from the discovery of $\Xi_{cc}^{++}(3621)$
\cite{LHCb:2017iph} and showing that the upper
bound on a ${\cal T}_{cc}^+$ mass
is approximately 60~MeV above the $DD^*$ threshold. On the contrary,
the bound on a ${\cal T}_{bb}$ mass was 224~MeV below the threshold.
In the same paper, we advocated the possibility of using the chiral
quark soliton model ($\chi$QSM) to estimate the ${\cal T}_{QQ}$ mass.

A mean field description of heavy
baryons as a light quark-soliton and a heavy quark has been introduced and
developed
in Refs.~\cite{Yang:2016qdz,Kim:2017jpx,Kim:2017khv,Polyakov:2022eub}.
This approach is a modi{\-}fication of the $\chi$QSM used previously to describe
light baryons (see \cite{Diakonov:1987ty} and Refs.~\cite{Christov:1995vm,Alkofer:1994ph,Petrov:2016vvl}
for review) where the soliton is constructed from $N_c$ light quarks. To describe heavy
baryons one has to remove one light quark from the valence level and add a heavy quark instead.
In the large $N_c$ limit this replacement hardly changes the mean fields of the soliton.

Support for such a treatment can be inferred from Ref.~\cite{Goeke:2005fs} where the authors
studied soliton behavior in the limit where the current quark masses are $m \rightarrow \infty$. Although
such a limit may at first sight be in contradiction with the chiral symmetry, which is the main theoretical 
basis of the model, it gave  very good phenomenological results when compared to lattice data at finite
$m_{\pi}$. At sufficiently large $m$, the soliton ceases to exist, and the correct heavy quark limit is achieved.

In the $\chi$QSM, the soliton mass is given as a sum over the  energies of the valence quarks and the sea quark
energies computed with respect to the vacuum and appropriately regularized~\cite{Goeke:2005fs},
\begin{equation}
M_{\rm sol}=N_c\Big[ E_{\rm val}+\sum_{E_n<0} (E_n-E_n^{(0)})\Big].
\label{eq:Msol1}
\end{equation}
In the present context, Eq.~(\ref{eq:Msol1}) takes the following form:
\begin{align}
M_{\rm sol}&=(N_c-1)\Big[ E_{\rm val}+\sum_{E_n<0} \left(E_n-E_n^{(0)}\right)\Big]  \label{eq:Msol2} \\
                  &+\Big[ E_{\rm val}(m_Q)+\sum_{E_n<0} \left(E_n(m_Q)-E_n^{(0)}(m_Q)\right)\Big].\notag
\end{align}
As  has been argued in Ref.~\cite{Goeke:2005fs}, for large $m_Q$,  the sum
over the sea quarks in the second
line of Eq.~(\ref{eq:Msol2}) vanishes, and $E_{\rm val}(m_Q)\approx m_Q$. One copy of the
soliton ceases to exist; however, the remaining $N_c-1$ quarks still form a stable soliton.

Such a soliton does not carry
any quantum numbers except for the baryon number resulting from the
valence quarks. Spin and isospin appear when the soliton rotations in space and flavor are quantized.
This procedure results in a {\em collective} Hamiltonian analogous to the one
of a quantum mechanical symmetric top; however, due to the Wess-Zumino-Witten
term \cite{WittenCA,Wess:1971yu},
the allowed Hilbert space is truncated to the representations that contain states of
hypercharge $Y'=N_{\rm val}/3$. For $N_{\rm val}=N_c=3$ (\ref{eq:Msol1}), these are octet and decuplet of ground state
baryons. For $N_{\rm val}=N_c-1=2$ (\ref{eq:Msol2}), we have antitriplet of spin 0 and sextet of spin 1. It is therefore
convenient to label heavy quark baryons (and tetraquarks as well) by the SU(3)$_{\rm flavor}$ representation
of the light subsystem.

From this perspective, the soliton is
reminiscent of a diquark, and the quantization rule $Y^{\prime}=(N_{c}-1)/3$
selects SU(3)$_{\mathrm{flavor}}$ representations identical as the ones of the
quark model. 
Given the success of the $\chi$QSM  in reproducing 
the data~\cite{Yang:2016qdz,Kim:2017jpx,Kim:2017khv,Polyakov:2022eub}, we
propose here to use the same strategy to describe the doubly heavy 
tetraquarks  replacing heavy quark $Q$ by a heavy (anti)diquark $\bar{Q}\bar{Q}$.

We observe that two heavy quarks of the same flavor (say $cc$ or $bb$) can form
a color antitriplet (antisymmetric in color) provided they are symmetric in
spin~\cite{Gelman:2002wf}. Therefore, they form a tight object of spin 1. Hence, two heavy antiquarks are
 in color $\mathbf{3}$ and spin 1, behaving as a spin 1
heavy {\em quark}. A tetraquark can be therefore viewed as being composed of a
heavy (anti)diquark of spin 1 and a ($N_{c}-1$)-quark soliton.\footnote{In what follows, we use
term {\em diquark} referring both to $QQ$ and $\bar{Q}\bar{Q}$ states.}

There are three main lessons that we have learned from our previous studies of
heavy baryons~\cite{Yang:2016qdz,Kim:2017jpx,Kim:2017khv,Polyakov:2022eub}:
\begin{itemize}
  \setlength\itemsep{0.01em}
 \item the soliton properties do not depend on the mass of the heavy quark,
\item neither do they depend on the spin coupling between a soliton and a heavy quark,
\item hyperfine splittings scale like $1/m_{Q}$.
\end{itemize}

This is discussed in detail in Sec.~\ref{sec:baryons}.

Therefore, a very simple and predictive picture of a soliton + heavy object (that
is $\boldsymbol{\overline{3}}$ in color) bound state emerges, where the mass
is simply given as a sum of the soliton mass (including $m_{s}$ and rotational
splittings), mass of a heavy object (quark or a diquark), and the hyperfine splitting.
This picture is very reminiscent to the one of Ref.~\cite{Eichten:2017ffp}. 
Mass formulas for such states are therefore identical to the ones of heavy baryons,
with some modification due to the spin 1 character of the heavy diquark; this is
elaborated on in detail in Sec.~\ref{sec:tetras}.
So the main problem is to estimate the diquark mass. Here, we propose to use
the Cornell potential as described later in Sec.~\ref{sec:diquark}.

We find that only bottom antitriplet tetraquarks, both nonstrange and strange, are bound by
approximately 140 and 60~MeV, respectively. We present numerical results for 
antitriplet and sextet tetraquarks in Sec.~\ref{sec:masses} and conclude in Sec.~\ref{sec:koniec}.

\section{Chiral Quark Soliton Model for Baryons}
\label{sec:baryons}

Let us first recall how baryon masses are calculated in the present model. We
quantize the soliton as if it were constructed from $N_{c}-1$ rather than $N_{c}$
light quarks. Then, in the chiral limit the soliton energy is given as%
\begin{equation}
E_{\text{sol}}=M_{\mathrm{sol}}+\frac{J(J+1)}{2I_{1}}+\frac{C_{2}%
(p,q)-J(J+1)-3/4\,Y^{\prime\,2}}{2I_{2}}.
\end{equation}
Here, $M_{\mathrm{sol}}$ is a classical soliton mass, $I_{1,2}$ denote moments
of inertia, $C_{2}(p,q)$ is the SU(3)$_{\rm flavor}$ Casimir, and $J$ corresponds to spin. In
our case $Y^{\prime}=2/3$ and the allowed SU(3) representations correspond to
$\boldsymbol{\overline{3}}$ with spin $J=0$ and $\boldsymbol{6}$ with spin $J=1$~\cite{Yang:2016qdz}.

SU(3) splittings are given by the operator%
\begin{equation}
H_{\mathrm{br}}=\alpha\,D_{88}^{(8)}+\beta\,\hat{Y}+\frac{\gamma}{\sqrt{3}%
}\sum_{i=1}^{3}D_{8i}^{(8)}\,\hat{J}_{i},
\end{equation}
where constants $\alpha$, $\beta$ and $\gamma$ can be expressed through
generalized moments of inertia (see e.g. Eq.~(4) in Ref~\cite{Yang:2016qdz})
and can be computed {\em ab initio} in some specific versions of the model.
In the most simple case with the pseudoscalar fields only, the numerical values
can be found e.g. in Ref.~\cite{Blotz:1992pw}, and in the context of heavy
baryons in Ref.~\cite{Kim:2019rcx}. In both cases they lead to reasonable
phenomenology. However, in reality one should take into account all possible
chiral fields: scalar, pseudoscalar, vector, axial and tensor \cite{Diakonov:2013qta}
for which full numerical analysis has not been performed. Here, the
explicit forms of  $\alpha,~\beta$, and $\gamma$ are not  needed
as we  treat them as free parameters. 

Heavy baryon masses are calculated by adding the mass of the heavy quark to
the soliton mass and by taking into account the hyperfine splitting given by
the following Hamiltonian:
\begin{equation}
H_{SQ}=\frac{2}{3}\frac{\varkappa}{m_{Q}} {\boldsymbol{J}}\cdot{\boldsymbol{S}}_{Q}
\label{eq:ssinter}%
\end{equation}
where ${\boldsymbol J}$ and ${\boldsymbol S}_Q$ stand for the soliton and heavy quark spin,
respectively. We have assumed here that the possible $m_{Q}$ dependence of
$\varkappa$, due to the presence of the wave function squared in
(\ref{eq:ssinter}), can be ignored. Since the spin of the
$\boldsymbol{\overline{3}}$ representation is zero, there is no hyperfine
splitting in this case. In the case of $\boldsymbol{6}$ we have two sets of
heavy baryons with spins $1/2$ and $3/2.$ This pattern is seen in the experimental data~\cite{Workman:2022ynf}.

Mass formulas for heavy baryons read therefore as follows~\cite{Praszalowicz:2019lje}:
\begin{align}
M_{B,\overline{\boldsymbol{3}}}=m_Q+M_{\mathrm{sol}}&+\frac{1}{2I_{2}}%
+\delta_{\boldsymbol{\overline{3}}}Y_{B} \notag \\
M_{B,\boldsymbol{6},s}    =m_Q+M_{\mathrm{sol}}&+\frac{1}{2I_{2}}
+%
\frac{1}{I_{1}}+\delta_{\boldsymbol{6}}Y_{B} \notag \\
& +\frac{\varkappa}{m_{Q}}\left\{
\begin{array}
[c]{ccc}%
-2/3 & \text{for} & s=1/2\\
\, & \, & \,\\
+1/3 & \text{for} & s=3/2
\end{array}
\right. \, .
\label{eq:M3barM6mass}%
\end{align}
Here splitting parameters $\delta_{\boldsymbol{\overline{3}}}$ and
$\delta_{\boldsymbol{6}}$ are known functions of parameters $\alpha$,
$\beta$ and $\gamma$ (see Eq.~(9) in Ref.~\cite{Yang:2016qdz}), and
$Y_B$ stands for a hypercharge of a given baryon.

Let us examine the consequences of the mass formulas (\ref{eq:M3barM6mass}).
First of all, as  has been already observed in \cite{Yang:2016qdz}, Eqs.~(\ref{eq:M3barM6mass})
admit one parameter independent sum rule in the sextet
\begin{align}
M_{\Omega_{Q}^{\ast}} 
= 
 2M_{\Xi^{\prime}_{Q}}
 +M_{\Sigma_{Q}^{\ast}} 
 -2 M_{\Sigma_{Q}} \, ,
\label{eq:OMpred}
\end{align}
which for charm is satisfied at the level $1.4$~MeV. We use (\ref{eq:OMpred}) 
to estimate $M_{\Omega_{b}^{\ast}}=6076.37$~MeV
when we compute average sextet masses in the $b$ sector.

To get rid of the hyperfine splittings, we average out spin dependence in sextets
by defining
\begin{equation}
M_{B,\boldsymbol{6}} =\frac{1}{3}\left( M_{B,\boldsymbol{6},1/2} +2\, M_{B,\boldsymbol{6},3/2} \right).
\label{eq:MB6ave}
\end{equation}
Average masses $M_{B,\boldsymbol{6}} $ and masses in triplets should be equally spaced with $Y_B$.
independently of the heavy quark.\footnote{We neglect small isospin violation.} 
For  $\bar{\bf 3}$ we have (in
MeV)
\begin{align}
-\delta_{\overline{\boldsymbol 3}}
=\left. 182.6 
\right|_{\Xi_c-\Lambda_c}
=\left. 174.9  
\right|_{\Xi_b-\Lambda_b}, 
\label{eq:deltabar3}  
\end{align}
which is satisfied with 2~\% accuracy. In the case of   $\boldsymbol{6}$, we 
have more relations (in MeV),
\begin{align}
 -\delta_{\boldsymbol{6}}& 
= \left. 126.7   \right|_{\Xi_c-\Sigma_c}
= \left. 119.1   \right|_{\Omega_c - \Xi_c} 
\cr
& 
= \left. 121.5 \right|_{\Xi_b-\Sigma_b}
= \left. 118.4  \right|_{\Omega_b - \Xi^{\ast}_b}
 \label{eq:delta6}
\end{align}
which are satisfied with 4~\% accuracy.\footnote{Note that for numerical analysis in the present paper,
we have used most recent version of PDG~\cite{Workman:2022ynf}, and therefore, there are small numerical differences with respect
to Ref.~\cite{Yang:2016qdz}.}

We can also form differences of average multiplet masses between the $b$ and $c$ sectors to compute
the heavy quark mass difference (in MeV),
\begin{equation}
m_b-m_c=\left. 3328\right|_{\overline{\boldsymbol{3}}}=\left. 3327\right|_{{\boldsymbol{6}}}\, .
\label{eq:mQdiff}
\end{equation}

Furthermore, we can extract the hyperfine splitting parameter testing our assumptions concerning
the Hamiltonian (\ref{eq:ssinter}),
\begin{align}
\frac{\varkappa}{m_c}
&= \left. 64.6\right|_{\Sigma_c}
 = \left. 67.2\right|_{\Xi_c} 
 = \left. 70.7\right|_{\Omega_c} \, ,
\cr
\frac{\varkappa}{m_b}
&= \left. 19.4 \right|_{\Sigma_b}
 = \left. 18.8\right|_{\Xi_b}
\label{eq:spintest}
\end{align}
(in MeV). From these estimates, we get
\begin{equation}
\frac{m_c}{m_b}\simeq 0.27 \div 0.30
\label{eq:mQratio}
\end{equation}
with the average value of $0.283$. The PDG values of the $\overline{\rm MS}$ heavy quark masses  
lead to $m_c/m_b=0.3$  where both masses
are evaluated at the renormalization point  $\mu=m_Q$~\cite{Workman:2022ynf}.
Of course, heavy quark masses in the effective models, like the one considered in this
paper, may differ from the QCD masses. It is therefore encouraging that we get a mass ratio
close to the ratio of the QCD masses. Nevertheless, quark masses extracted from Eqs.~(\ref{eq:mQdiff})
and (\ref{eq:mQratio}),
\begin{align}
m_{c}  &  =1206\div1426\;\text{MeV,}\nonumber\\
m_{b}  &  =4533\div 4753\;\text{MeV} 
\label{eq:mQbaryon}%
\end{align}
are a bit higher (especially for $m_b$) than those quoted by PDG~\cite{Workman:2022ynf}.
For $m_c/m_b=0.283$, we get $m_c=1314.1$~MeV and $m_b=4641.5$~MeV, which are
still lower than the effective values  used, e.g., in Ref.~\cite{Karliner:2014gca}.

Finally, to test heavy quark dependence of the mass formulas (\ref{eq:M3barM6mass}),
we can compute the nonstrange  moment of inertia from the sextet-${\overline{\boldsymbol 3}}$
average mass differences,
\begin{align}
\frac{1}{I_1}=M_{\mathbf{6}}^Q -
  M_{\mathbf{\overline{3}}}^Q 
=\left. 171.5\right|_c=\left. 170.4\right|_b 
\label{eq:I1Q}
\end{align}
in MeV. We see that indeed heavy quark masses cancel with very high precision. This, together
with Eq.~(\ref{eq:mQdiff}), suggests
that possible nonlinear in $m_Q$ binding effects are very small if not vanishing. We can therefore safely assume
that formulas (\ref{eq:mQdiff}) are valid for any heavy object replacing $Q$. We pursue this possibility
in the next section.

%%%%%%%%%%%%%%%%%%%%%%%%%%%%%%%%%%%%%%%%%%%%%%%%%%%%%%%%

\section{Chiral Quark Soliton Model for Tetraquarks}
\label{sec:tetras}

In the present case, instead of a heavy quark, we add to the soliton a heavy
diquark $\bar{Q}\bar{Q}$ of spin 1. Assuming that the soliton
is not changed by this replacement we arrive at the following mass formulas
for tetraquarks:%
\begin{align}
M_{Q}^{\text{tetra}\,\boldsymbol{\overline{3}}}  &  =M_{B,\overline{\boldsymbol{3}}}
+m_{\bar{Q}\bar{Q}}-m_{Q},\nonumber\\
M_{Q}^{\text{tetra}\,\boldsymbol{6}}  &  =M_{B,\boldsymbol{6}}+m_{\bar{Q}\bar{Q}}-m_{Q}
+C_{s}\frac{2}{3}\frac{\varkappa}{m_Q}\frac{m_Q}{m_{\bar{Q}\bar{Q}}}%
\label{eq:allmasses}
\end{align}
where $C_{s}$ is a spin factor arising from the fact that both the sextet soliton
and the diquark have spin 1:%
\begin{equation}
C_{s}=\left\{
\begin{array}
[c]{rcc}%
-2 & \text{for} & s=0\\
-1 & \text{for} & s=1\\
1 & \text{for} & s=2
\end{array}
\right.  .
\label{eq:tetramass}
\end{equation}
Here $M_{B,\overline{\boldsymbol{3}}}$ is a heavy baryon mass in SU(3)$_{\rm flavor}$
$\overline{{\boldsymbol 3}}$ and $M_{B,\boldsymbol{6}}$ is a spin averaged mass
of a sextet baryon (\ref{eq:MB6ave}).

Sextet splittings satisfy the following relation
\begin{eqnarray}
\Delta^{\boldsymbol{6}}_{\rm spin}&=&(M_{Q}^{\text{tetra}\,\boldsymbol{6}}(s=1)-M_{Q}^{\text{tetra}\,\boldsymbol{6}}(s=0)) \notag \\
&=&\frac{1}{2}(M_{Q}^{\text{tetra}\,\boldsymbol{6}}(s=2)-M_{Q}^{\text{tetra}\,\boldsymbol{6}}(s=1))  \notag \\
&=& \frac{2}{3}\frac{\varkappa}{m_Q} \frac{m_Q}{m_{QQ}}
\label{eq:6Delta}
\end{eqnarray}

Before proceeding to numerical calculations let us  discuss strong decay thresholds.
Since the ground state $\boldsymbol{\overline{3}}$ tetraquarks have
$J^{P}=1^{+}$, they can decay to $D+D^{\ast}$ or $B+B^{\ast}$. The corresponding
thresholds are listed in the second rows of Tables~\ref{tab:th} and \ref{tab:ths}
for nonstrange and strange tetraquarks, respectively. In the latter case, $D_s D^*$ and $B_s B^*$ thresholds
are lighter than $D D_s^{\ast}$ or $B B_s^{\ast}$.

In the case of the sextet tetraquarks, we have three
families of spin $0$, 1, and $2$ of nonstrange, strange and doubly strange tetraquarks.
Pertinent thresholds (averaged over isospin) are listed in Tables~\ref{tab:th}--\ref{tab:thss}.
\renewcommand{\arraystretch}{1.2} 

\begin{table}[h]
\centering
\begin{tabular}[c]{|c|c|c|}%
\hline
$J^{P}$ & {Channel} & {Thresholds [MeV]}\\
\hline
$0^{+}$ & $DD,BB$ & 3736.1\quad10558.9\\
$1^{+}$ & $DD^{\ast},BB^{\ast}$ & 3877.2\quad10604.2\\
$2^{+}$ & $D^{\ast}D^{\ast},B^{\ast}B^{\ast}$ & 4018.3\quad10649.4\\
\hline
\end{tabular}
\caption{Thresholds for nonstrange tetraquarks decays. \label{tab:th}}
\end{table}

\begin{table}[h]
\centering
\begin{tabular}[c]{|c|c|c|}%
\hline
$J^{P}$ & {Channel} & {Thresholds [MeV]}\\
\hline
$0^{+}$ &$ D_{s}D,B_{s}B$ & 3836.4\quad10646.4\\
$1^{+} $& $D_{s}D^{\ast},B_{s}B^{\ast}$ & 3977.5\quad10691.6\\
$2^{+}$ &$ D_{s}^{\ast}D^{\ast},B_{s}^{\ast}B^{\ast}$ & 4121.3\quad10740.1\\
\hline
\end{tabular}
\caption{Thresholds for strange tetraquarks decays. \label{tab:ths}}
\end{table}

\begin{table}[h]
\centering
\begin{tabular}[c]{|c|c|c|}%
\hline
$J^{P}$ & {Channel} & {Thresholds [MeV]}\\
\hline
$0^{+}$ & $D_{s}D_{s},B_{s}B_{s} $& 3936.7\quad10733.8\\
$1^{+}$ & $ D_{s}D_{s}^{\ast},B_{s}B_{s}^{\ast}$ & 4080.6\quad10782.3\\
$2^{+}$ & $D_{s}^{\ast}D_{s}^{\ast},B_{s}^{\ast}B_{s}^{\ast}$ & 4224.4\quad10830.8 \\
\hline
\end{tabular}
\caption{Thresholds for doubly strange tetraquarks decays. \label{tab:thss}}
\end{table}

Mass formulas (\ref{eq:tetramass}) relate tetraquark masses directly to heavy
baryon masses and therefore are fairly model independent. They are analogous 
to the masses given in Eq.(1) of Ref.~\cite{Eichten:2017ffp}. The spin  part has been
discussed in \cite{Eichten:2017ffp} and in \cite{Karliner:2021wju}; however, the hyperfine coupling has not
been specified. Here, we know the value of
$\varkappa/m_{c,b}$ (\ref{eq:spintest}),  so in order to estimate  tetraquark
masses, we only need  heavy diquark mass $m_{\bar{Q}\bar{Q}}$
for $m_Q$ in the range (\ref{eq:mQbaryon}).

%%%%%%%%%%%%%%%%%%%%%%%%%%%%%%

\section{Heavy Diquark Mass}
\label{sec:diquark}

The main problem in predicting heavy tetraquark masses in the present model is
to have a reliable estimate of the heavy diquark mass, as it is beyond the large $N_c$
effective theory that we have used for the light sector. To this end, we propose to
apply a nonrelativistic Schr{\"o}dinger equation with the Cornell potential~\cite{Eichten:1978tg}
\begin{equation}
V(r)=-\frac{\kappa }{r}+\sigma \,r \, 
\label{eq:Cornell}
\end{equation}
with $\kappa=C_F \alpha_{\rm s}$, which has been successfully used to describe
heavy $Q\bar{Q}$ spectra (see, e.g., Ref.~\cite{Mateu:2018zym}). 

\begin{figure}[h]
\centering
\includegraphics[width=7.cm]{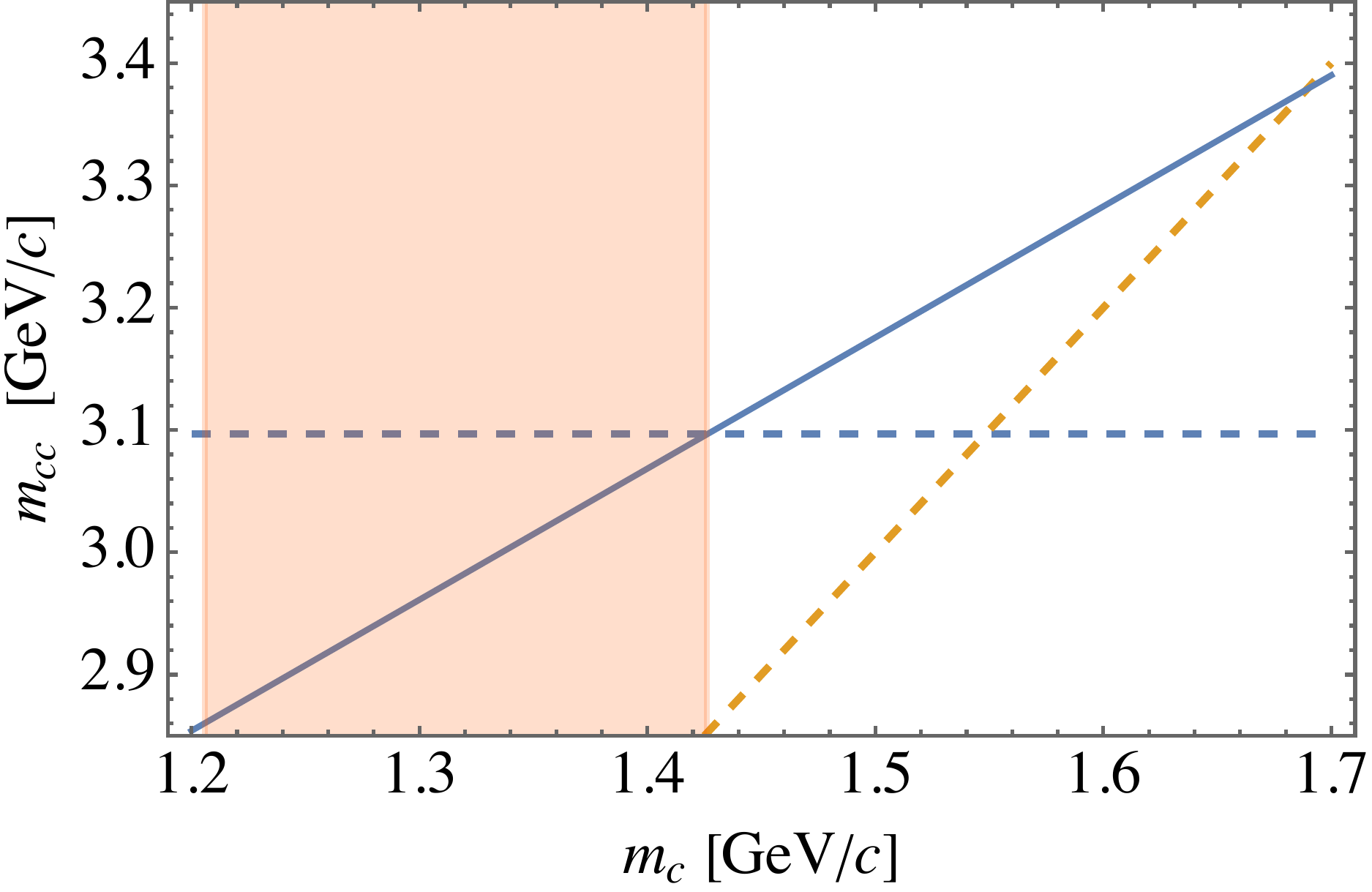} \\
\includegraphics[width=7.cm]{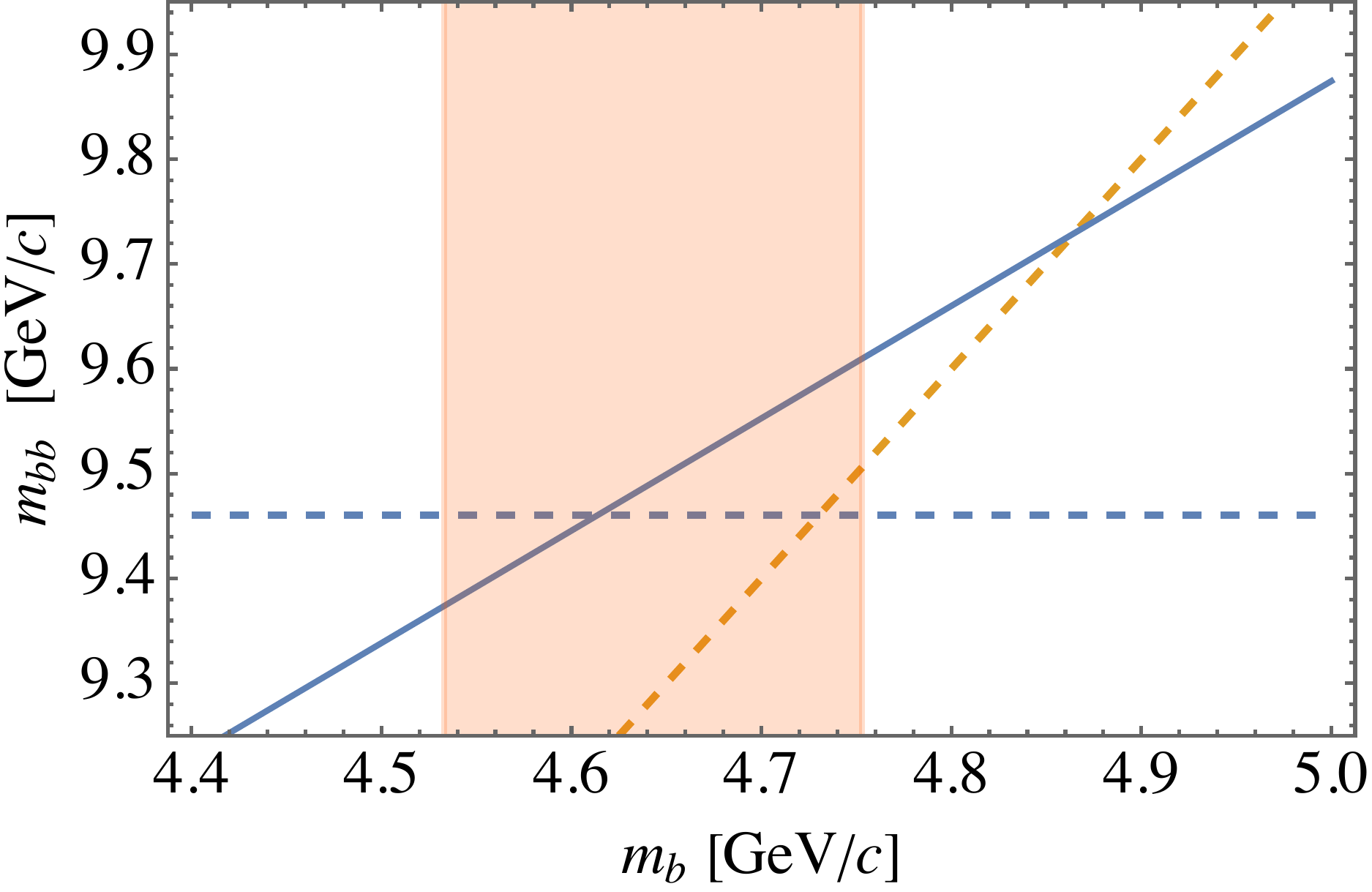} 
\caption{Diquark masses from the Cornell potential (\ref{eq:Cornell}) as functions of $m_Q$ (solid).
Horizontal dashed blue lines correspond to $J/\psi$ or $\Upsilon$ for charm and bottom, respectively.
Oblique orange dashed lines show $2m_Q$. Shaded areas indicate the heavy
quark mass ranges (\ref{eq:mQbaryon}) deduced from the heavy baryon spectra.}%
\label{fig:mQQmass}%
\end{figure}

There are two
practical reasons to use the Cornell potential in the present context. The first one is
that in order to compute $QQ$ (or $\bar{Q}\bar{Q}$) masses one has to rescale model
parameters by a factor of 2. This follows from the fact that the color charge 
$\left< {\boldsymbol \lambda}\cdot {\boldsymbol \lambda}\right>$ is factor 2 smaller
when quark color charges are in an (anti)triplet than in a singlet (see, e.g., Table III in Ref.~\cite{Karliner:2014gca}). 
As this is quite obvious for the
Coulomb term, lattice calculations suggest the same behavior of the confining
part \cite{Nakamura:2005hk}. Also the chromomagnetic spin interaction, which we neglect
in the following, scales in the same way.

The second reason is that the Coulomb part in potential (\ref{eq:Cornell}) can be in fact considered
as a perturbation to the linear potential, for which solutions in terms of the Airy functions
are known semianalytically. We have checked that it is enough to consider only the first order perturbation theory. 

We are interested in the $S$ states only, so we put $l=0$ in the pertinent Schr{\"o}dinger equation. The reduced mass of the
equal mass system entering the Schr{\"o}dinger equation is $\mu=m_Q/2$. So we are looking for a solution in terms of a $u_n$ function
defined as follows:
\begin{equation}
\psi_{nlm}(r,\theta ,\varphi )=R_{0}^{n}(r)Y_{00}(\theta ,\varphi )= \frac{u_{n}(r)}{r} \frac{1}{\sqrt{4\pi}}.
\end{equation}
It is convenient to introduce a dimensionless variable  $\rho$, 
\begin{equation}
r=\left( \frac{\hbar ^{2}}{\sigma m_Q }\right) ^{1/3} \rho
\end{equation}
and rescaled dimensionless parameters $\lambda$ and $\zeta$,
\begin{eqnarray}
\lambda =\left( \frac{m_Q}{\sigma
^{1/2}\hbar ^{2}}\right) ^{2/3}\kappa ,  
&~~~&
%\notag \\
\zeta =\left( \frac{m_Q}{\sigma ^{2}\hbar ^{2}}%
\right) ^{1/3}E.  \label{eq:newpars}
\end{eqnarray}
With these substitutions, the Schr{\"o}dinger equation takes a very simple form,
\begin{equation}
 u^{\prime \prime }+\left[ \frac{\lambda }{\rho }-\rho
+\zeta \right] u =0.  \label{eq:fullu}
\end{equation}

For $\lambda=0$, Eq.~(\ref{eq:fullu}) reduces to the Airy equation, and the unperturbed energies are given in
terms of the zeros $z_n$ of the Airy function  ${\rm Ai}(\rho-\zeta)$. This follows from the boundary condition $u_n(0)=0$.
Therefore, we have energy quantization,%
\begin{equation}
\zeta _{n}^{(0)}=-z_{n} \, .
\end{equation}%
Note that these zeros are
negative, so the energy $\zeta _{n}^{(0)}$ is positive. The normalized
solution is%
\begin{equation}
u_{n}(\rho )=\mathcal{N}_{n}\, {\rm Ai}(\rho -\zeta _{n}^{(0)})=\mathcal{N}%
_{n}\, {\rm Ai}(\rho +z_{n}).
\end{equation}

First order perturbative correction is linear in $\lambda$, so the full energy reads
\begin{equation}
\zeta _{n}=-z_{n}-\lambda a_{n} \, .
\label{eq:zetasol}
\end{equation}
We  need energies for two first levels only, for which $a_1=0.835$ and $a_2=0.582$.
Masses of the $Q\bar{Q}$ states read
\begin{align}
M_{n}=&2m_Q+\left( \frac{\sigma ^{2}\hbar ^{2}}{m_Q}\right) ^{1/3}\left(
-z_{n}-\lambda a_{n} \right) \notag \\
=&2m_{Q}-\varepsilon _{Q}z_{n}-\frac{\tilde{\kappa}}{\varepsilon _{Q}}a_{n} 
\label{eq:2mass}
\end{align}%
where we have introduced two new parameters,
\begin{equation}
\varepsilon _{Q}=\left( \frac{\sigma ^{2}\hbar ^{2}}{m_{Q}}\right)^{1/3}~~~~{\rm and}~~~~~\tilde{\kappa}=\kappa\sigma=\varepsilon_Q^2\lambda \, .
\end{equation}

For a given $m_Q$ from  the range covering (\ref{eq:mQbaryon}), we have computed  parameters $\varepsilon_Q$
and $\tilde{\kappa}$ from the two lowest $Q\bar{Q}$ states.\footnote{Parameters $\varepsilon_Q$
and $\tilde{\kappa}$ must be positive. It turns out that there are no such solutions for too low $m_Q$.}
Since we need to estimate the mass of a
spin 1 diquark, we have chosen as inputs $J/\psi(3096.6)$ and $\psi_{2S}(3686.1)$ for charm
and $\Upsilon_{1S}(9399.0)$ and $\Upsilon_{2S}(10023.3)$ for bottom. We have checked that
the original parameters $\kappa$ and $\sigma$ obtained that way are in qualitative agreement with numerical
results of Ref.~\cite{Mateu:2018zym}.

Having $\varepsilon_Q$ and $\tilde{\kappa}$ fixed, we can easily compute diquark masses in color (anti)triplet
by rescaling $\kappa \rightarrow \kappa/2$ and $\sigma \rightarrow \sigma/2$, leading to 
$\varepsilon_Q \rightarrow \varepsilon_Q/4^{1/3}$
and  $\tilde{\kappa} \rightarrow \tilde{\kappa}/4$. It is important to realize that the two terms in Eq.~(\ref{eq:2mass})
scale differently with this change of parameters. The confining positive part is reduced by a factor $(1/4)^{1/3} \simeq 0.63$
while the Coulomb negative part is reduced  by  $(1/4)^2/3 \simeq 0.4$. This delicate balance can make the
diquark mass higher than the $Q\bar{Q}$ ground state. This happens, however, only at sufficiently high $m_Q$
where the first order perturbation theory breaks down.

The
diquark masses for charm and bottom are plotted in Fig.~\ref{fig:mQQmass}. One can see that at sufficiently large mass
the Coulomb term becomes equal to the confining term, and the diquark mass becomes lighter than $2m_Q$ signaling
the break down of the first order perturbation theory.  However, in the range of model masses (\ref{eq:mQbaryon}), the linear
confining term dominates, and the first order perturbation theory is sufficient. In Ref.~\cite{Praszalowicz:2019lje}, we have naively
approximated $m_{QQ}\simeq 2m_Q$, whereas for the Cornell potential, we get  $m_{QQ}\simeq (2.1 \div 2.3)\,m_Q$
in the mass range (\ref{eq:mQbaryon}). This seemingly small difference led to the overbinding observed in \cite{Praszalowicz:2019lje}.

It is of course legitimate to ask how the diquark masses  depend on the potential that one chooses
do describe heavy quark dynamics. One could try, for example, a harmonic oscillator potential, which for $m_c=1400$~MeV
and $\omega=590$~MeV reproduces masses of $J/\psi$ and $\psi_{2S}$. After rescaling $\omega^2 \rightarrow \omega^2/2$
(which is a naive implementation of the rescaling valid for the Cornell potential),
one obtains $m_{cc}=3008$~MeV, 60~MeV below the mass following from the Cornell potential for $m_c=1400$~MeV.
Nevertheless, as we shall shortly see, this reduction does not lead to a bound tetraquark state.

\section{Tetraquark Masses}
\label{sec:masses}
\subsection{Antitriplet masses}

It is now straightforward to compute predictions for the tetraquarks in
flavor $\boldsymbol{\overline{3}}$ with the help of Eqs.(\ref{eq:allmasses})
and the numerical results for the diquark masses from the previous section.
The results are plotted in Fig.~\ref{fig:3barmass} and listed in Table~\ref{tab:m3bar}. We can see that charm
tetraquark masses are above the threshold, while in the case of bottom
we see rather deeply bound states both for nonstrange and strange tetraquarks.
The lightest nonstrange charm tetraquark is approximately 70~MeV above the 
$DD^*$ threshold, so even the harmonic oscillator model for the heavy diquark
would not lead to binding. Our results are in a very good agreement with
predictions of Ref.~\cite{Eichten:2017ffp}, although up to 30~MeV lower.

\renewcommand{\arraystretch}{1.4} 

\begin{table}
\begin{tabular}[c]{|c|c|c|}
\hline
& Charm & Bottom\\
\hline
$m_{Q}$ & 1.31 & 4.64\\
\hline
$T_{QQq_1q_2}^{\overline{\boldsymbol{3}}} $& 3.95 & 10.47\\
$T_{QQsq}^{\overline{\boldsymbol{3}}}$ & 4.13 & 10.64 \\
\hline
\end{tabular}
\caption{Masses of antitriplet tetraquarks in GeV. \label{tab:m3bar}}
\end{table}

\begin{widetext}
\begin{center}
\begin{figure}[h]
\centering
\includegraphics[height=5.5cm]{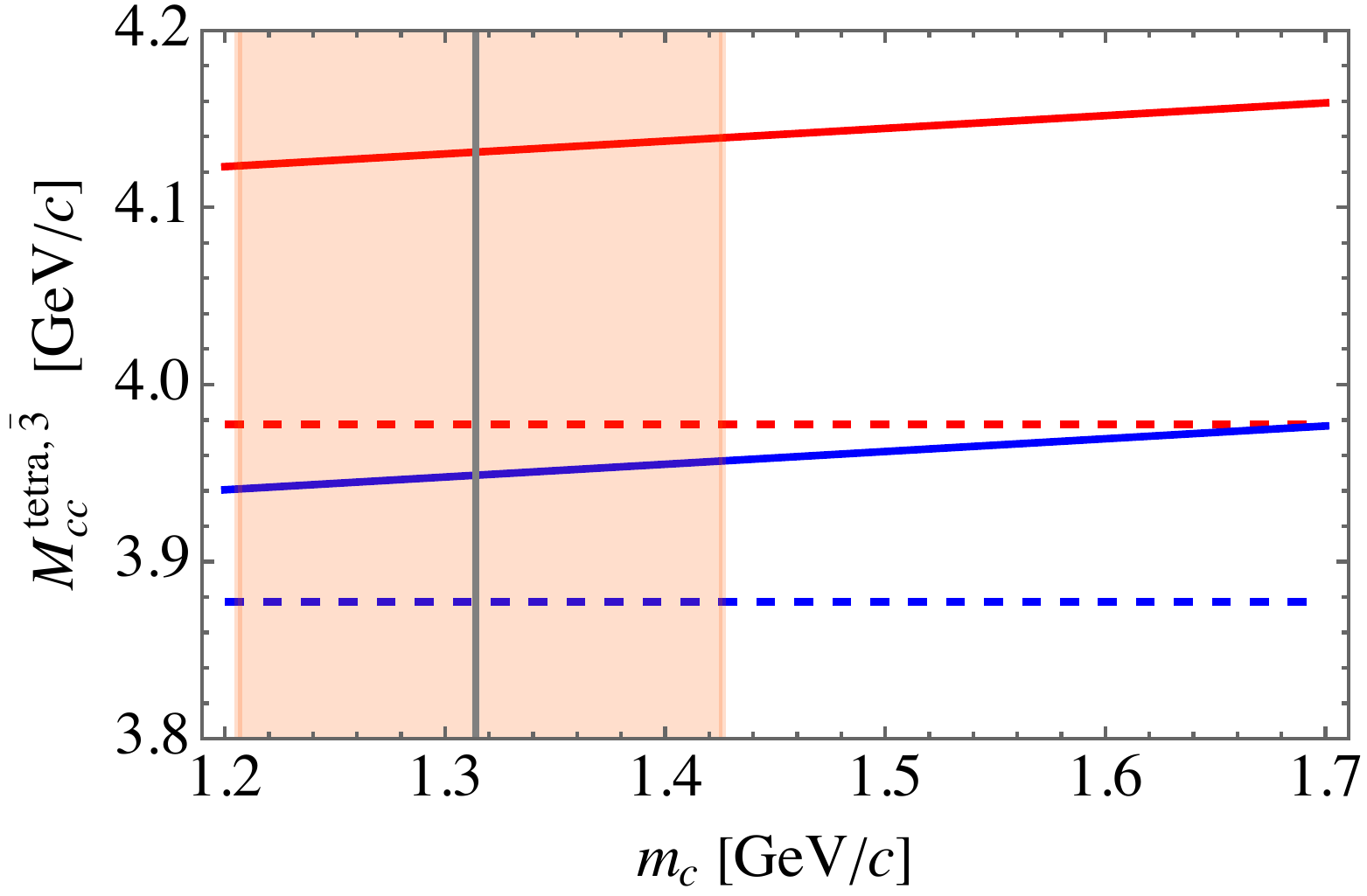}~~~~~~~
\includegraphics[height=5.5cm]{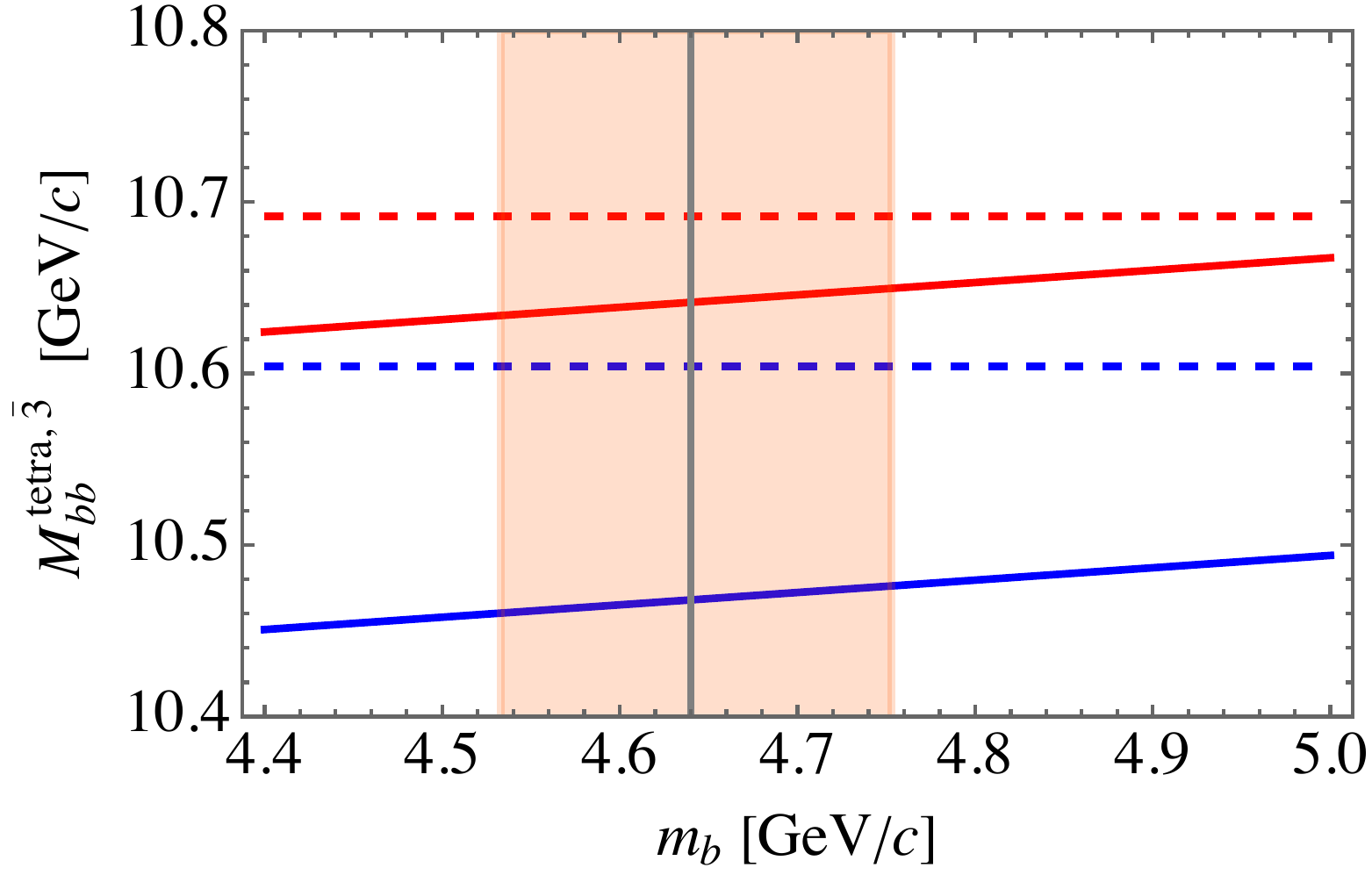} 
\caption{The lightest
nonstrange (solid blue, bottom) and strange (solid red, top) antitriplet tetraquark masses (charm, left panel; bottom, right panel)
as  functions of the heavy quark mass. Horizontal dashed lines correspond to the pertinent thresholds
(nonstrange, bottom;  strange, top)
discussed in Sec.~\ref{sec:tetras}. Shaded areas indicate the heavy
quark mass range (\ref{eq:mQbaryon}). Solid vertical
lines correspond to $m_c=1314$~MeV or $m_b=4641.5$~MeV.}%
\label{fig:3barmass}%
\end{figure}
\end{center}
%\begin{widetext}
\begin{center}
\begin{figure}[h!]
\centering
\includegraphics[height=4.8cm]{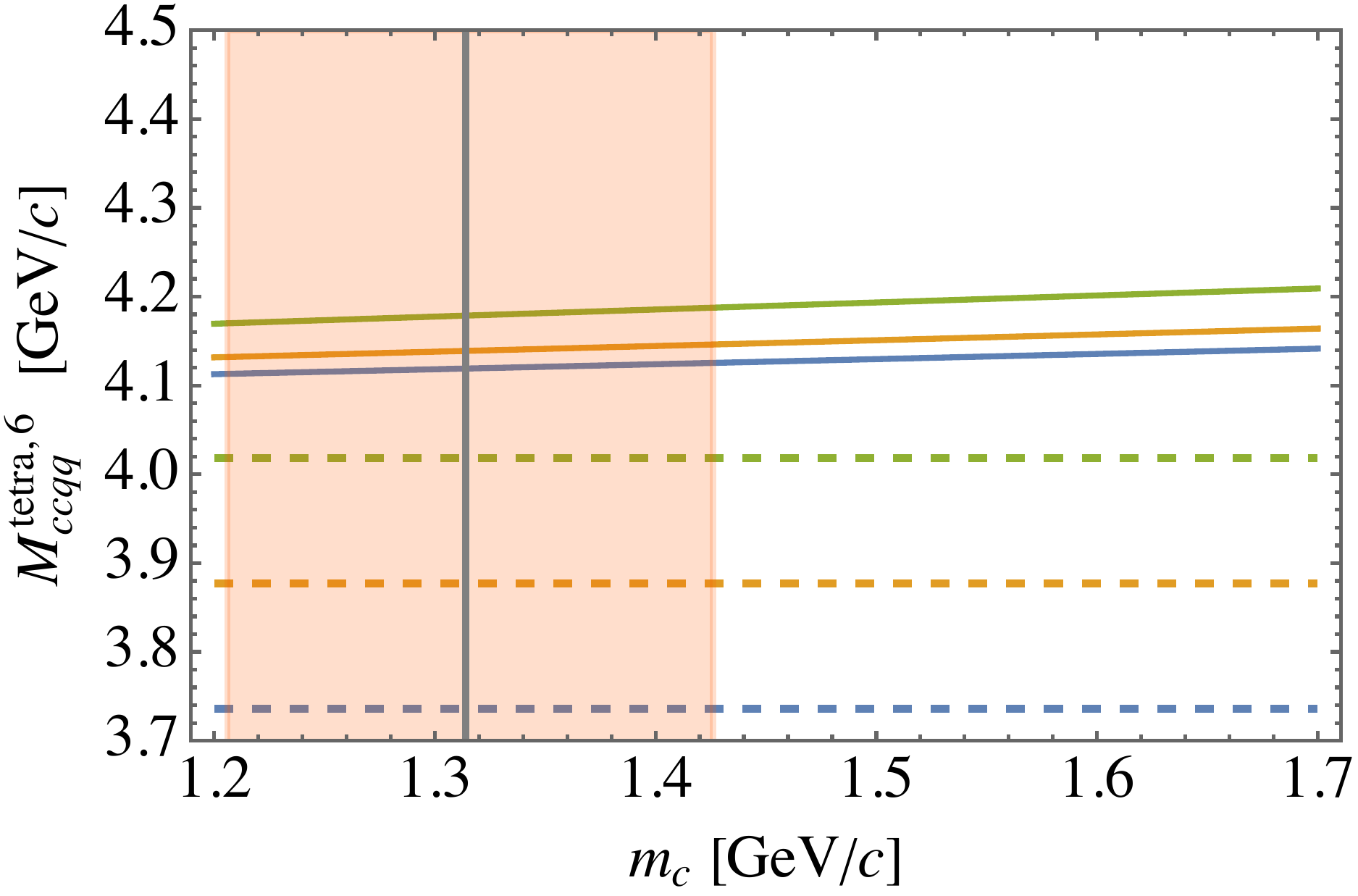}~~~~~~\includegraphics[height=4.8cm]{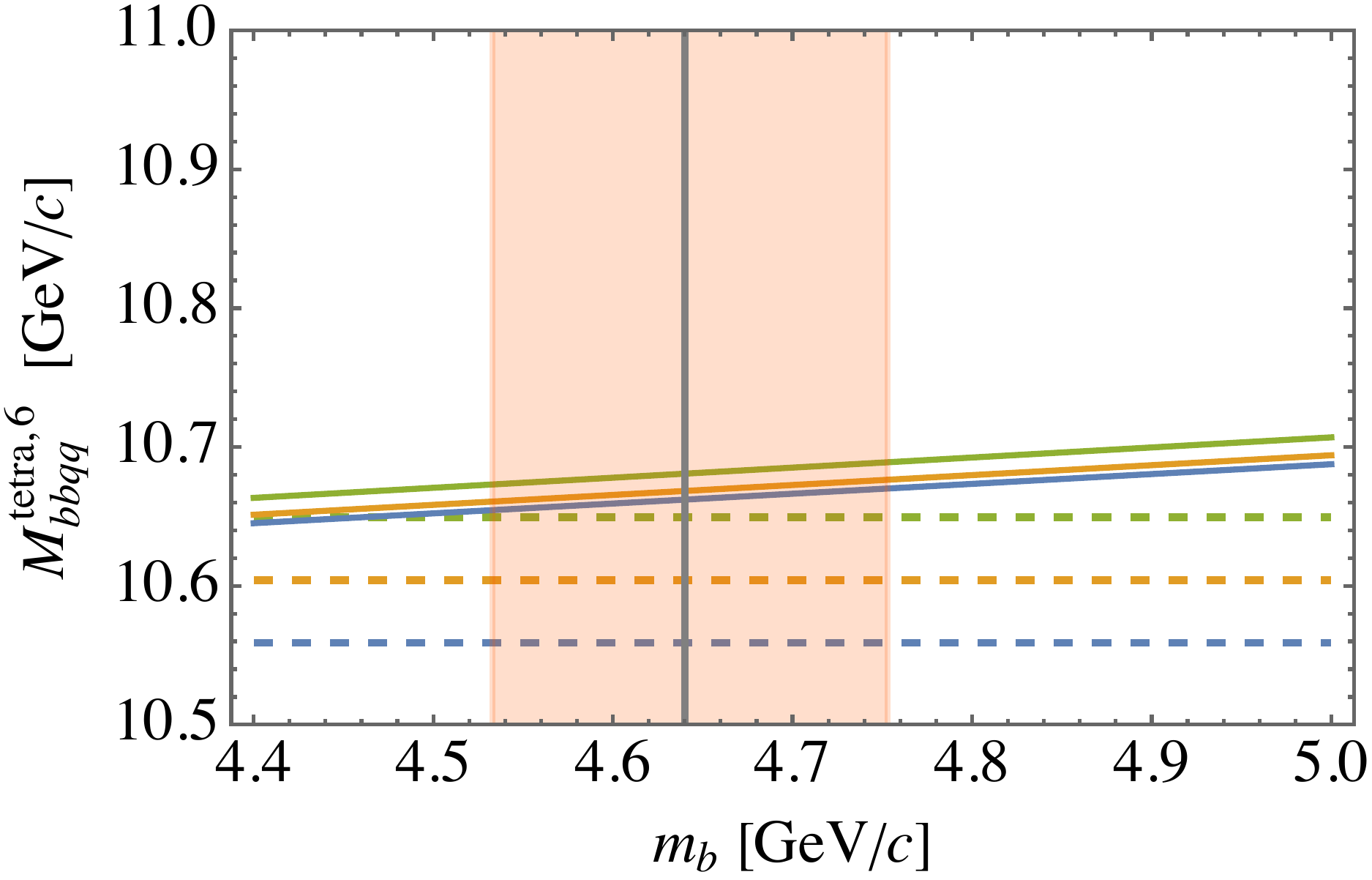}\\
\includegraphics[height=4.8cm]{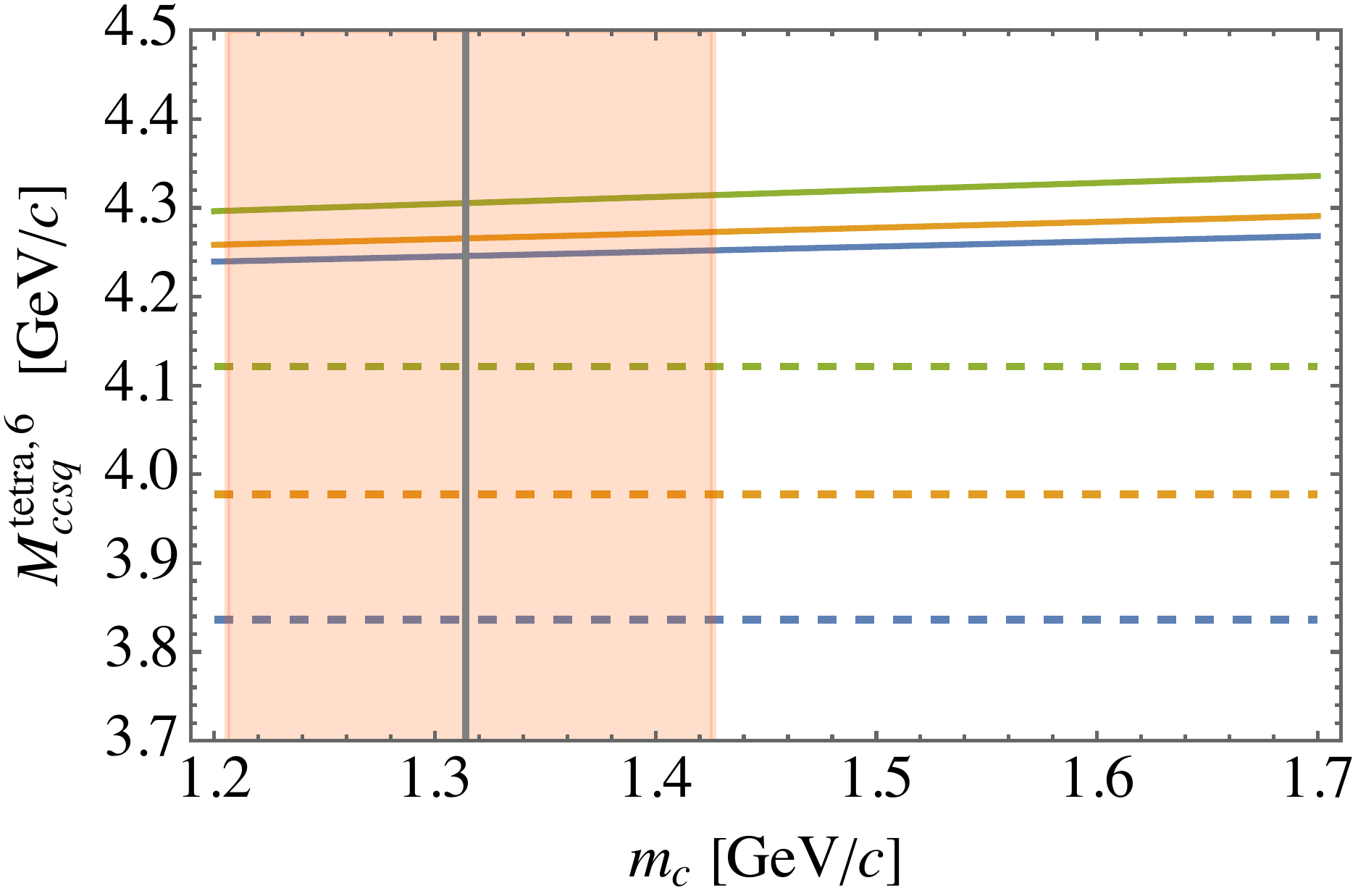}~~~~~~\includegraphics[height=4.8cm]{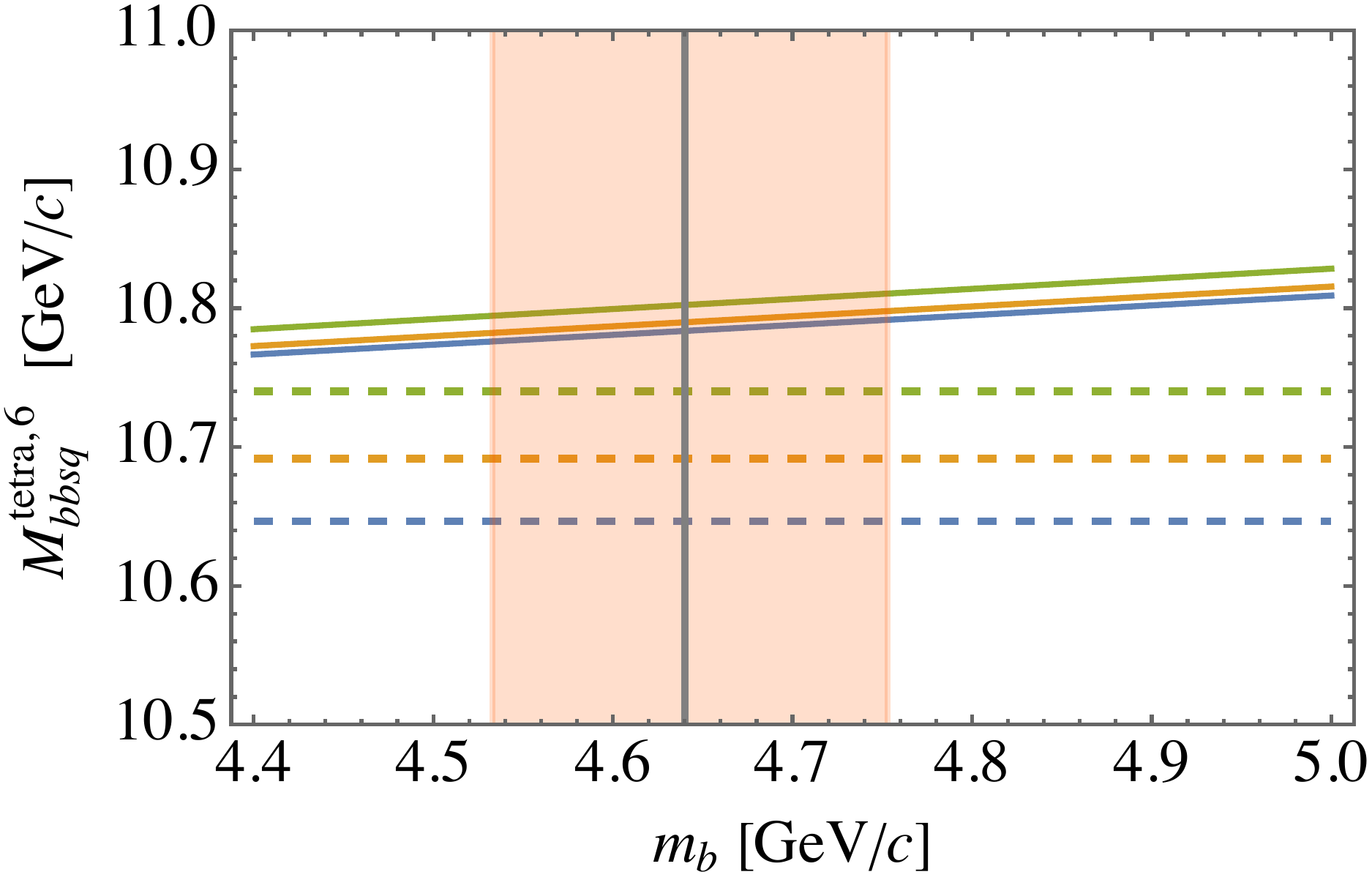} \\
\includegraphics[height=4.8cm]{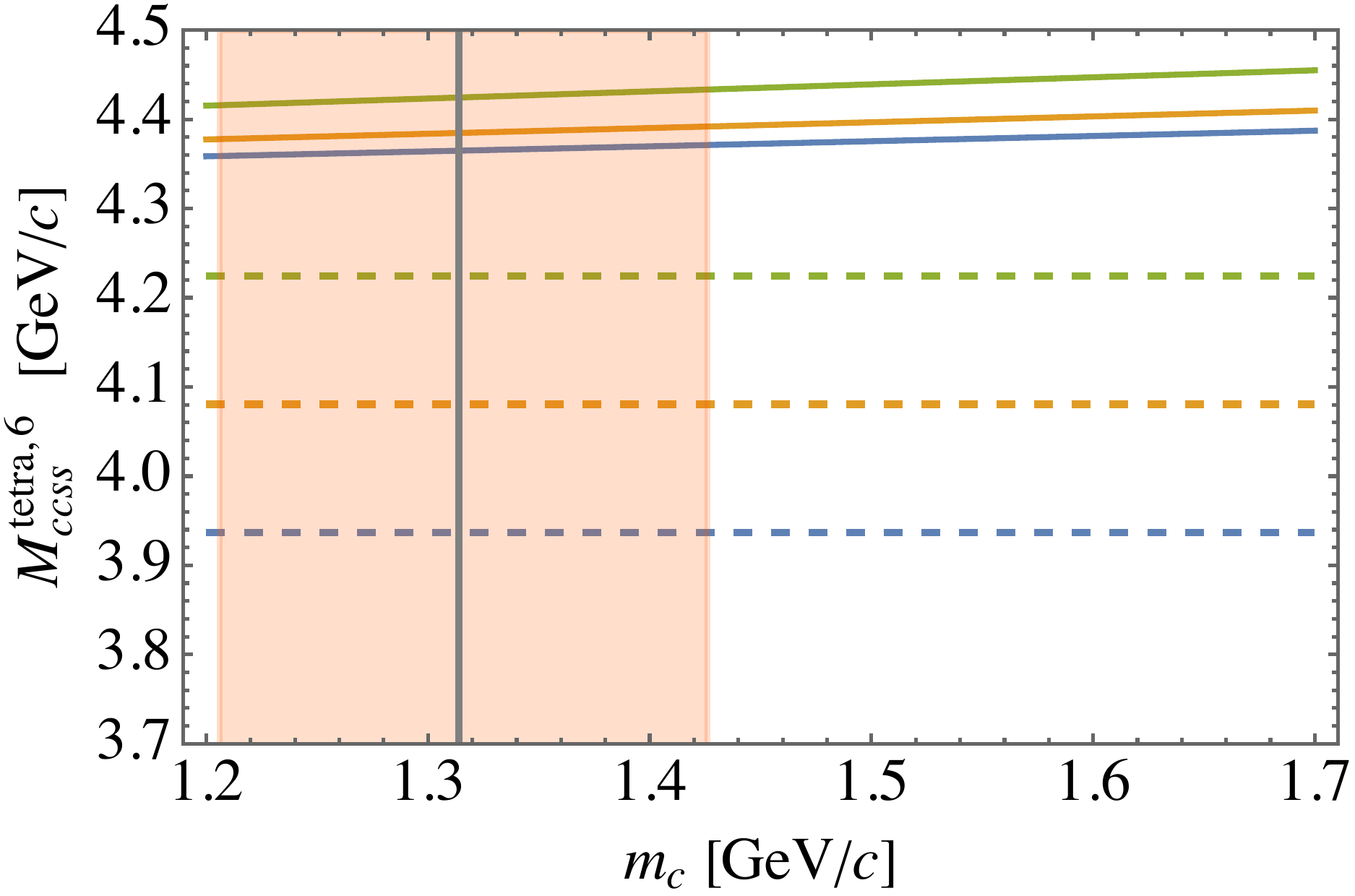}~~~~~~\includegraphics[height=4.8cm]{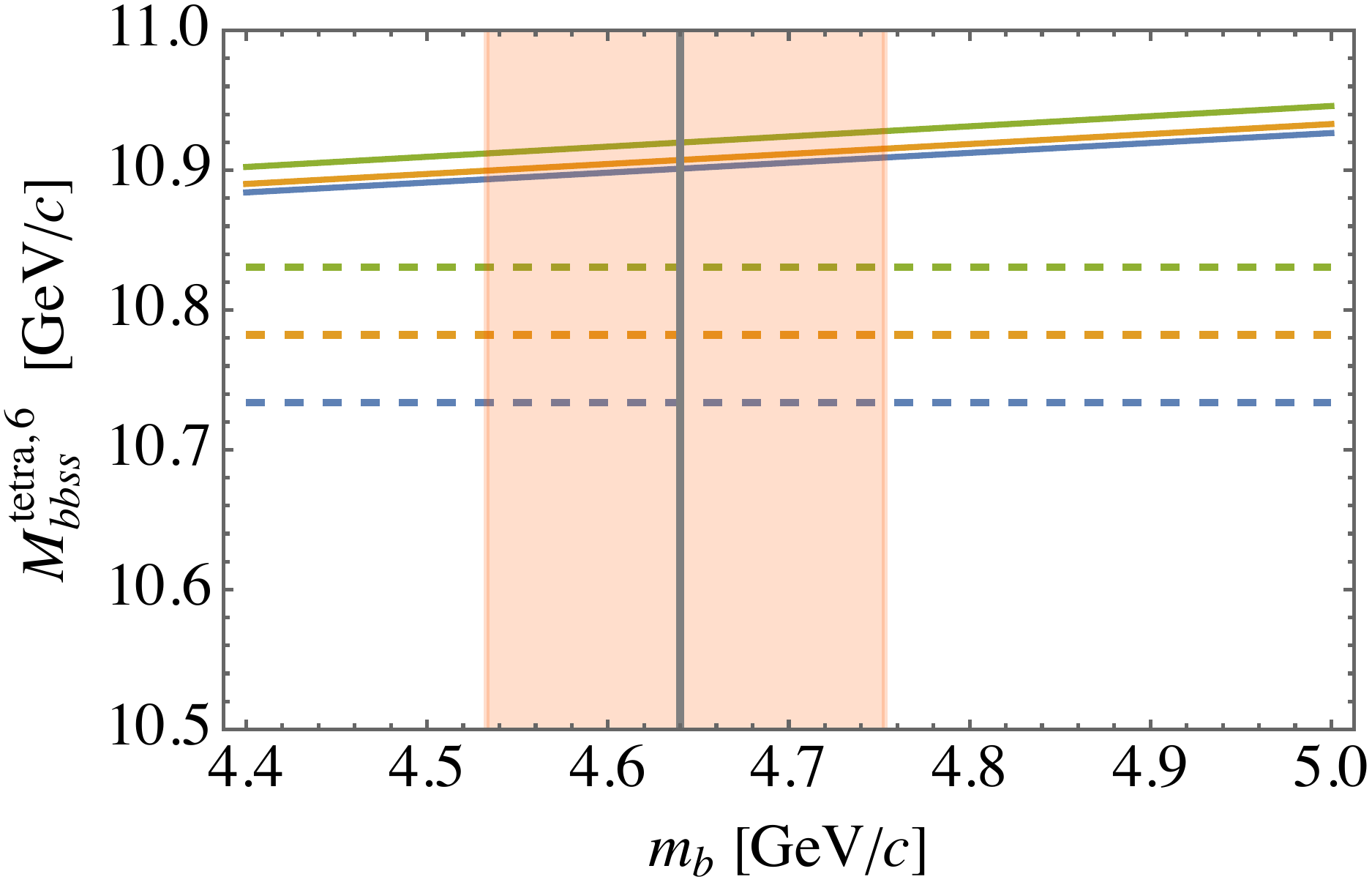} 
\caption{The lightest
nonstrange, strange,  and doubly strange  sextet tetraquark masses (charm, left; bottom, right)
of spin 0 (solid blue, bottom), spin 1 (solid orange, middle) and spin 2 (solid green, top)
as  functions of the heavy quark mass. Horizontal dashed lines correspond to the pertinent thresholds
(in the same order from bottom to top as the masses) shown in Tables~~\ref{tab:th}, \ref{tab:ths}, and \ref{tab:thss}.
%discussed in Sec.~\ref{sec:tetras}. 
Shaded areas indicate the heavy
quark mass range (\ref{eq:mQbaryon}). Solid vertical
lines correspond to $m_c=1314$~MeV or $m_b=4641.5$~MeV.}%
\label{fig:6mass}%
\end{figure}
\end{center}
\end{widetext}

\subsection{Sextet  masses}

The only difference between the antitriplet masses and sextet masses is the
presence of the hyperfine splitting. Interestingly, from (\ref{eq:6Delta}),
we expect for charm $\Delta^{\boldsymbol{6}}_{\rm spin} \simeq 19 \div 21$~MeV, as $m_Q/m_{QQ}$
in the range (\ref{eq:mQbaryon}) is approximately $0.43 \div 0.48$. On the contrary, hyperfine
splitting $D^*-D\simeq 140$~MeV is 7 times larger (and similarly in the $b$ sector). So in fact different spin states
in the sextet are almost degenerate.
We see this clearly in Fig.~\ref{fig:6mass} where we plot predictions for the sextet tetraquark masses (solid lines) 
and the pertinent thresholds (dashed lines). Different colors correspond to spin. 
%Upper panels are for nonstrange
%tetraquarks lower panels correspond to tetraquarks with one or two strange quarks. We see that none of the sextet tetraquarks,
%both charm or bottom, is bound. 
The only possible candidate for a bound state, given the accuracy of the present model, is
a nonstrange bottom tetraquark of spin 2, which is only $\sim 30$~MeV above the threshold. Numerical values can be
found in Table~\ref{tab:m6}.

%\newpage

\section{Summary and Conclusions}
\label{sec:koniec}

Motivated by the success of the chiral quark soliton model in describing the heavy baryon spectra, we have constructed
mass formulas for heavy tetraquarks with two heavy quarks of the same flavor. We first discussed baryon phenomenology 
to conclude that the properties of the light sector do not depend on the heavy quark properties. This is quite expected on the
grounds of heavy quark symmetry. It is therefore legitimate to replace heavy quark $Q$ in color $\boldsymbol{3}$ by a heavy
anti-diquark, that differs from $Q$ by mass and spin. Mass formulas (\ref{eq:M3barM6mass})  relate tetraquark masses  
to the masses of heavy baryons, and the only model parameter borrowed from the baryon phenomenology is the hyperfine
splitting parameter (\ref{eq:spintest}) $\varkappa/m_Q$. In this sense, our approach, although derived from the $\chi$QSM,
is fairly model independent. This is why formulas (\ref{eq:M3barM6mass}) are identical to the ones derived in the heavy quark
limit from QCD in Ref.~\cite{Eichten:2017ffp}.

\begin{table}
\begin{tabular}[c]{|c|c|c|c|c|c|c|}
\hline
 & \multicolumn{3}{c|}{Charm}  &\multicolumn{3}{c|}{Bottom} \\
 \hline
$m_{Q}$ & \multicolumn{3}{c|}{1.31}  &\multicolumn{3}{c|}{4.64} \\
\hline
$s$ & 0 & 1 & 2 & 0 & 1 & 2\\
\hline
$T_{QQq_1q_2}^{\boldsymbol{6}} $& 4.12 & 4.14 & 4.18 & 10.66 & 10.67 & 10.68\\
$T_{QQsq}^{\boldsymbol{6}}$ & 4.25 & 4.27 & 4.31 & 10.78 & 10.79 & 10.80\\
$T_{QQss}^{\boldsymbol{6}}$ & 4.37 & 4.38 & 4.42 & 10.90 & 10.91 & 10.92 \\
\hline
\end{tabular}
\caption{Masses of sextet tetraquarks in GeV. \label{tab:m6}}
\end{table}

The only unknown ingredient of the present approach is the heavy diquark mass. To this end, we have used the Cornell
potential, first to fit potential parameters to reproduce lowest spin 1 onia, both in charm and bottom sectors, and then, 
after rescaling these parameters, to compute the spin 1 diquark masses. We find that only bottom tetraquarks in flavor
antitriplet are bound, while the charm ones are above the threshold. This is true also in the case when the structure of a heavy diquark
can be resolved by the light quarks and repulsive color ${\bf 6}$ channel is included \cite{Czarnecki:2017vco}.

Numerical results presented in Tables \ref{tab:m3bar} and \ref{tab:m6} are in a very good agreement with the results
of Ref.~\cite{Eichten:2017ffp} where all necessary parameters have been extracted from  data, including the
mass of $\Xi_{cc}^{++}$~\cite{LHCb:2017iph}. No model calculations have been performed in Ref.~\cite{Eichten:2017ffp},
and in turn, we did not use any input from doubly charmed $\Xi$.
This is a strong argument in favor of our approach to the heavy diquark mass.

\begin{figure}[h]
\centering
\includegraphics[width=6.0cm]{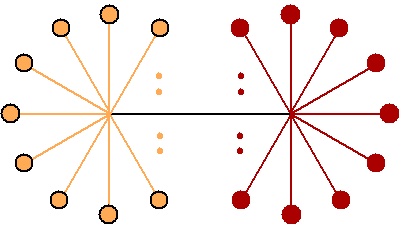}
\caption{Heavy tetraquark at large $N_c$. Full circles denote light quarks; full circles
with black contours are for antiquarks.}%
\label{fig:NcTetra}%
\end{figure}

It is interesting to observe that our model has a completely different $N_c$ counting than {\em typical models} discussed, { e.g.},
in Refs.~\cite{Gelman:2002wf} or \cite{Czarnecki:2017vco}, where tetraquarks are composed from four quarks for any  $N_c$.
In our case, the soliton for large $N_c$ belongs to a color representation $\cal{R}$ corresponding to an antisymmetric product on $N_c-1$
quarks. This is because we have to take one light quark from the soliton and add one heavy quark to construct
a heavy baryon. For $N_c=2$, this is $\cal{R}={\overline{\boldsymbol{ 3}}}$. In order to construct a tetraquark, we need to put
heavy antiquarks  in a complex
conjugate representation $\overline{\cal{R}}$ corresponding to $N_c-1$ antisymmetrized antiquarks. 
For $N_c=2$, this is ${\boldsymbol 3}$. So for arbitrary
$N_c$, our tetraquark consits of $N_c-1$ light quarks and $N_c-1$ heavy antiquarks, see Fig.~\ref{fig:NcTetra}.
Such a configuration has been briefly discussed in Ref.~\cite{Gelman:2002wf}. 
A system composed of $N_c-1$ heavy (anti)quarks is amenable to semiclassical treatment. It would be interesting
to pursue this possibility in constructing a model for a diquark.

Finally, we have to confront the LHCb tetraquark \cite{LHCb:2021vvq,LHCb:2021auc} which is just below the $DD^*$ threshold. Here, two
possibilities exist. Either our model is not accurate enough to deal with dynamics which gives binding energies of the order of
hundreds keV, or the LHCb tetraquark corresponds to a different configuration that is out of reach for the soliton models. Obviously,
charm quark mass is far from infinity and $1/m_c$ corrections might finally lower our predictions. 
In the present approach, however, we have no systematic scheme that would allow one to include such effects.
Also the diquark model can be responsible
for overshooting the physical mass. Nevertheless, given the very good accuracy of the $\chi$QSM predictions for heavy baryon masses
and very good agreement with the phenomenological analysis of Ref.~\cite{Eichten:2017ffp}, one is
perhaps more inclined towards the second possibility. Indeed, the LHCb~\cite{LHCb:2021auc} estimated the size of ${\cal T}_{cc}^+$ 
to be of the
order of $7$~fm, significantly larger than the typical size of heavy flavor hadrons. 
This suggests a molecular structure of ${\cal T}_{cc}^+$~\cite{Janc:2004qn,Li:2012ss}.

In order to compute the space (or momentum) structure of tetraquarks in the present model, one should resort to a dynamical
description of the soliton in terms of quark degrees of freedom. Some studies in this direction 
within $\chi$QSM
have been undertaken 
in the case of singly heavy baryons. In Ref.~\cite{Kim:2018nqf} electromagnetic form factors 
and in Refs.~\cite{Kim:2020nug,Won:2022cyy} gravitational form factors have been studied. 
It follows that heavy baryons are more compact than the proton. That conclusion should also apply to the present case,
as the heavy quark or diquark is treated here merely as a static color source. The internal structure
of heavy tetraquarks certainly deserves detailed studies, it is, however, beyond scope of the present paper.

\bigskip

\section*{Acknowledgements}
This research has been 
supported by the Polish National Science Centre Grants No. 2017/27/B/ST2/01314
and No. 2018/31/B/ST2/01022.
The author thanks the Institute for Nuclear Theory at the University of Washington for its kind hospitality, stimulating research environment
and partial support
by the INT's U.S. Department of Energy Grant No. DE-FG02-00ER41132.


\begin{thebibliography}{10}
%\cite{LHCb:2021vvq}
\bibitem{LHCb:2021vvq}
R.~Aaij \textit{et al.} [LHCb],
%``Observation of an exotic narrow doubly charmed tetraquark,''
Nature Phys. \textbf{18}, no.7, 751-754 (2022)
doi:10.1038/s41567-022-01614-y
[arXiv:2109.01038 [hep-ex]].
%121 citations counted in INSPIRE as of 10 Aug 2022

%\cite{LHCb:2021auc}
\bibitem{LHCb:2021auc}
R.~Aaij \textit{et al.} [LHCb],
%``Study of the doubly charmed tetraquark $T_{cc}^{+}$,''
Nature Commun. \textbf{13}, no.1, 3351 (2022)
doi:10.1038/s41467-022-30206-w
[arXiv:2109.01056 [hep-ex]].
%115 citations counted in INSPIRE as of 10 Aug 2022

%\cite{Karliner:2017qhf}
\bibitem{Karliner:2017qhf}
M.~Karliner, J.~L.~Rosner and T.~Skwarnicki,
%``Multiquark States,''
Ann. Rev. Nucl. Part. Sci. \textbf{68}, 17-44 (2018)
doi:10.1146/annurev-nucl-101917-020902
[arXiv:1711.10626 [hep-ph]].
%167 citations counted in INSPIRE as of 10 Aug 2022

%\cite{Chen:2022asf}
\bibitem{Chen:2022asf}
H.~X.~Chen, W.~Chen, X.~Liu, Y.~R.~Liu and S.~L.~Zhu,
%``An updated review of the new hadron states,''
[arXiv:2204.02649 [hep-ph]].
%44 citations counted in INSPIRE as of 10 Aug 2022

%\cite{Carlson:1987hh}
\bibitem{Carlson:1987hh}
J.~Carlson, L.~Heller and J.~A.~Tjon,
%``Stability of Dimesons,''
Phys. Rev. D \textbf{37}, 744 (1988)
doi:10.1103/PhysRevD.37.744
%138 citations counted in INSPIRE as of 16 Aug 2022

%\cite{Manohar:1992nd}
\bibitem{Manohar:1992nd}
  A.~V.~Manohar and M.~B.~Wise,
  %``Exotic Q Q anti-q anti-q states in QCD,''
 {\em  Nucl.\ Phys.\  B} {\bf 399}, 17 (1993).
  %doi:10.1016/0550-3213(93)90614-U
 % [hep-ph/9212236].
  %%CITATION = doi:10.1016/0550-3213(93)90614-U;%%
  %177 citations counted in INSPIRE as of 09 Mar 2019
  
  %\cite{Isgur:1991wq}
\bibitem{Isgur:1991wq}
N.~Isgur and M.~B.~Wise,
%``Spectroscopy with heavy quark symmetry,''
Phys. Rev. Lett. \textbf{66}, 1130-1133 (1991)
doi:10.1103/PhysRevLett.66.1130
%586 citations counted in INSPIRE as of 11 Mar 2022

  
  %\cite{Cohen:2006jg}
\bibitem{Cohen:2006jg} 
  T.~D.~Cohen and P.~M.~Hohler,
  %``Doubly heavy hadrons and the domain of validity of doubly heavy diquark-anti-quark symmetry,''
  {\em Phys.\ Rev.\  D} {\bf 74}, 094003 (2006).
%  doi:10.1103/PhysRevD.74.094003
%  [hep-ph/0606084].
  %%CITATION = doi:10.1103/PhysRevD.74.094003;%%
  %27 citations counted in INSPIRE as of 09 Mar 2019
  
%\cite{Cai:2019orb}
\bibitem{Cai:2019orb}
Y.~Cai and T.~Cohen,
%``Narrow exotic hadrons in the heavy quark limit of QCD,''
Eur. Phys. J. A \textbf{55}, no.11, 206 (2019)
doi:10.1140/epja/i2019-12906-0
[arXiv:1901.05473 [hep-ph]].
%5 citations counted in INSPIRE as of 10 Aug 2022

%\cite{Eichten:2017ffp}
\bibitem{Eichten:2017ffp}
E.~J.~Eichten and C.~Quigg,
%``Heavy-quark symmetry implies stable heavy tetraquark mesons $Q_iQ_j \bar q_k \bar q_l$,''
Phys. Rev. Lett. \textbf{119}, no.20, 202002 (2017)
doi:10.1103/PhysRevLett.119.202002
[arXiv:1707.09575 [hep-ph]].
%201 citations counted in INSPIRE as of 10 Aug 2022

%\cite{Lipkin:1986dw}
\bibitem{Lipkin:1986dw}  H.~J.~Lipkin,  
%``A Model Independent Approach To Multi - Quark Bound States,''
{\em Phys.\ Lett.\  B} {\bf 172}, 242 (1986).
% doi:10.1016/0370-2693(86)90843-9  
%%CITATION = doi:10.1016/0370-2693(86)90843-9;%%
%108 citations counted in INSPIRE as of 20 Jan 2019

%\cite{Ader:1981db}
\bibitem{Ader:1981db}
  J.~P.~Ader, J.~M.~Richard and P.~Taxil,
  %``Do Narrow Heavy Multi - Quark States Exist?,''
  {\em Phys.\ Rev.\ D} {\bf 25},  2370 (1982).
%  doi:10.1103/PhysRevD.25.2370
  %%CITATION = doi:10.1103/PhysRevD.25.2370;%%
  %153 citations counted in INSPIRE as of 05 Apr 2019
  
  %\cite{Praszalowicz:2019lje}
\bibitem{Praszalowicz:2019lje}
M.~Prasza{\l}owicz,
%``Doubly heavy $QQ$ tetraquarks,''
Acta Phys. Polon. Supp. \textbf{13}, 103-108 (2020)
doi:10.5506/APhysPolBSupp.13.103
[arXiv:1904.02676 [hep-ph]].
%1 citations counted in INSPIRE as of 10 Aug 2022


  
 %\cite{LHCb:2017iph}
\bibitem{LHCb:2017iph}
R.~Aaij \textit{et al.} [LHCb],
%``Observation of the doubly charmed baryon $\Xi_{cc}^{++}$,''
Phys. Rev. Lett. \textbf{119}, no.11, 112001 (2017)
doi:10.1103/PhysRevLett.119.112001
[arXiv:1707.01621 [hep-ex]].
%453 citations counted in INSPIRE as of 13 Aug 2022


%\cite{Yang:2016qdz}
\bibitem{Yang:2016qdz}
G.~S.~Yang, H.~C.~Kim, M.~V.~Polyakov and M.~Prasza\l{}owicz,
%``Pion mean fields and heavy baryons,''
Phys. Rev. D \textbf{94}, 071502 (2016)
doi:10.1103/PhysRevD.94.071502
[arXiv:1607.07089 [hep-ph]].
%34 citations counted in INSPIRE as of 10 Aug 2022

%\cite{Kim:2017jpx}
\bibitem{Kim:2017jpx}
H.~C.~Kim, M.~V.~Polyakov and M.~Prasza\l{}owicz,
%``Possibility of the existence of charmed exotica,''
Phys. Rev. D \textbf{96}, no.1, 014009 (2017)
doi:10.1103/PhysRevD.96.014009
[arXiv:1704.04082 [hep-ph]].
%68 citations counted in INSPIRE as of 10 Aug 2022

%\cite{Kim:2017khv}
\bibitem{Kim:2017khv}
H.~C.~Kim, M.~V.~Polyakov, M.~Prasza{\l}owicz and G.~S.~Yang,
%``Strong decays of exotic and nonexotic heavy baryons in the chiral quark-soliton model,''
Phys. Rev. D \textbf{96}, no.9, 094021 (2017)
[erratum: Phys. Rev. D \textbf{97}, no.3, 039901 (2018)]
doi:10.1103/PhysRevD.96.094021
[arXiv:1709.04927 [hep-ph]].
%36 citations counted in INSPIRE as of 10 Aug 2022

%\cite{Polyakov:2022eub}
\bibitem{Polyakov:2022eub}
M.~V.~Polyakov and M.~Prasza{\l}owicz,
%``Landscape of heavy baryons from the perspective of the chiral quark-soliton model,''
Phys. Rev. D \textbf{105}, 094004 (2022)
doi:10.1103/PhysRevD.105.094004
[arXiv:2201.07293 [hep-ph]].
%2 citations counted in INSPIRE as of 10 Aug 2022

%\cite{Diakonov:1987ty}
\bibitem {Diakonov:1987ty} D.~Diakonov, V.~Y.~Petrov and P.~V.~Pobylitsa,
%``A Chiral Theory of Nucleons,''
Nucl.\ Phys.\ B \textbf{306} (1988) 809.
%doi:10.1016/0550-3213(88)90443-9.
%%CITATION = doi:10.1016/0550-3213(88)90443-9;%%
%377 citations counted in INSPIRE as of 15 Jun 2016


%\cite{Christov:1995vm}


\bibitem {Christov:1995vm} C.~V.~Christov, A.~Blotz, H.~C.~Kim, P.~Pobylitsa,
T.~Watabe, T.~Meissner, E.~Ruiz Arriola and K.~Goeke,
%``Baryons as nontopological chiral solitons,''
Prog.\ Part.\ Nucl.\ Phys.\ \textbf{37} (1996) 91.
%doi:10.1016/0146-6410(96)00057-9
%[hep-ph/9604441].
%%CITATION = doi:10.1016/0146-6410(96)00057-9;%%
%275 citations counted in INSPIRE as of 31 Aug 2017


%\cite{Alkofer:1994ph}


\bibitem {Alkofer:1994ph} R.~Alkofer, H.~Reinhardt and H.~Weigel,
%``Baryons as chiral solitons in the Nambu-Jona-Lasinio model,''
Phys.\ Rept.\ \textbf{265} (1996) 139.
%doi:10.1016/0370-1573(95)00018-6
%[hep-ph/9501213].
%%CITATION = doi:10.1016/0370-1573(95)00018-6;%%
%223 citations counted in INSPIRE as of 17 May 2018


%\cite{Petrov:2016vvl}


\bibitem {Petrov:2016vvl} V.~Petrov,
%``Soliton Model for Baryons,''
Acta Phys.\ Polon.\ B \textbf{47} (2016) 59.
%doi:10.5506/APhysPolB.47.59
%%CITATION = doi:10.5506/APhysPolB.47.59;%%
%2 citations counted in INSPIRE as of 13 May 2018

\bibitem{Goeke:2005fs}
K.~Goeke, J.~Ossmann, P.~Schweitzer and A.~Silva,
%``Pion mass dependence of the nucleon mass and chiral
%extrapolation of lattice data in the chiral quark soliton model,''
Eur. Phys. J. A \textbf{27}, 77-90 (2006),
doi:10.1140/epja/i2005-10229-5,
[arXiv:hep-lat/0505010 [hep-lat]].


\bibitem{WittenCA}
%%CITATION = doi:10.1016/0550-3213(79)90232-3;%%
%2290 citations counted in INSPIRE as of 21 Jul 2016
E.~Witten,
%``Current Algebra, Baryons, and Quark Confinement,''
Nucl.\ Phys.\ B
\textbf{223} (1983) 422, and
%Nucl.\ Phys.\ B
\textbf{223} (1983) 433.
%doi:10.1016/0550-3213(83)90064-0.
%%CITATION = doi:10.1016/0550-3213(83)90064-0;%%
%1233 citations counted in INSPIRE as of 26 Aug 2016

%\cite{Wess:1971yu}
\bibitem{Wess:1971yu}
J.~Wess and B.~Zumino,
%``Consequences of anomalous Ward identities,''
Phys. Lett. B \textbf{37}, 95-97 (1971)
doi:10.1016/0370-2693(71)90582-X
%2898 citations counted in INSPIRE as of 02 Aug 2022

%\cite{Gelman:2002wf}
\bibitem{Gelman:2002wf}
B.~A.~Gelman and S.~Nussinov,
%``Does a narrow tetraquark cc anti-u anti-d state exist?,''
Phys. Lett. B \textbf{551}, 296-304 (2003)
doi:10.1016/S0370-2693(02)03069-1
[arXiv:hep-ph/0209095 [hep-ph]].
%87 citations counted in INSPIRE as of 10 Aug 2022

%\cite{Blotz:1992pw}
\bibitem{Blotz:1992pw}
A.~Blotz, D.~Diakonov, K.~Goeke, N.~W.~Park, V.~Petrov and P.~V.~Pobylitsa,
%``The SU(3) Nambu-Jona-Lasinio soliton in the collective quantization formulation,''
Nucl. Phys. A \textbf{555}, 765-792 (1993)
doi:10.1016/0375-9474(93)90505-R
%147 citations counted in INSPIRE as of 01 Nov 2022

%\cite{Kim:2019rcx}
\bibitem{Kim:2019rcx}
J.~Y.~Kim and H.~C.~Kim,
%``Improved pion mean fields and masses of singly heavy baryons,''
PTEP \textbf{2020}, no.4, 043D03 (2020)
doi:10.1093/ptep/ptaa037
[arXiv:1909.00123 [hep-ph]].
%10 citations counted in INSPIRE as of 01 Nov 2022

%\cite{Diakonov:2013qta}
\bibitem{Diakonov:2013qta}
D.~Diakonov, V.~Petrov and A.~A.~Vladimirov,
%``A theory of baryon resonances at large $N_c$,''
Phys. Rev. D \textbf{88}, no.7, 074030 (2013)
doi:10.1103/PhysRevD.88.074030
[arXiv:1308.0947 [hep-ph]].
%13 citations counted in INSPIRE as of 01 Nov 2022

%\cite{Workman:2022ynf}
\bibitem{Workman:2022ynf}
R.~L.~Workman \textit{et al.} [Particle Data Group],
%``Review of Particle Physics,''
PTEP \textbf{2022}, 083C01 (2022)
doi:10.1093/ptep/ptac097
%3 citations counted in INSPIRE as of 12 Jul 2022

%\cite{Karliner:2014gca}
\bibitem{Karliner:2014gca}
M.~Karliner and J.~L.~Rosner,
%``Baryons with two heavy quarks: Masses, production, decays, and detection,''
Phys. Rev. D \textbf{90}, no.9, 094007 (2014)
doi:10.1103/PhysRevD.90.094007
[arXiv:1408.5877 [hep-ph]].
%187 citations counted in INSPIRE as of 11 Aug 2022

%\cite{Karliner:2021wju}
\bibitem{Karliner:2021wju}
M.~Karliner and J.~L.~Rosner,
%``Doubly charmed strange tetraquark,''
Phys. Rev. D \textbf{105}, no.3, 034020 (2022)
doi:10.1103/PhysRevD.105.034020
[arXiv:2110.12054 [hep-ph]].
%5 citations counted in INSPIRE as of 11 Aug 2022

%\cite{Eichten:1978tg}
\bibitem{Eichten:1978tg}
E.~Eichten, K.~Gottfried, T.~Kinoshita, K.~D.~Lane and T.~M.~Yan,
%``Charmonium: The Model,''
Phys. Rev. D \textbf{17}, 3090 (1978)
[erratum: Phys. Rev. D \textbf{21}, 313 (1980)]
doi:10.1103/PhysRevD.17.3090
%1645 citations counted in INSPIRE as of 11 Aug 2022

%\cite{Mateu:2018zym}
\bibitem{Mateu:2018zym}
V.~Mateu, P.~G.~Ortega, D.~R.~Entem and F.~Fern\'andez,
%``Calibrating the Na\"\i{}ve Cornell Model with NRQCD,''
Eur. Phys. J. C \textbf{79}, no.4, 323 (2019)
doi:10.1140/epjc/s10052-019-6808-2
[arXiv:1811.01982 [hep-ph]].
%16 citations counted in INSPIRE as of 11 Aug 2022



%\cite{Nakamura:2005hk}
\bibitem{Nakamura:2005hk}
A.~Nakamura and T.~Saito,
%``QCD color interactions between two quarks,''
Phys. Lett. B \textbf{621}, 171-175 (2005)
doi:10.1016/j.physletb.2005.06.053
[arXiv:hep-lat/0512043 [hep-lat]].
%27 citations counted in INSPIRE as of 11 Aug 2022

%\cite{Czarnecki:2017vco}
\bibitem{Czarnecki:2017vco}
A.~Czarnecki, B.~Leng and M.~B.~Voloshin,
%``Stability of tetrons,''
Phys. Lett. B \textbf{778}, 233-238 (2018)
doi:10.1016/j.physletb.2018.01.034
[arXiv:1708.04594 [hep-ph]].
%59 citations counted in INSPIRE as of 12 Aug 2022

%\cite{Janc:2004qn}
\bibitem{Janc:2004qn}
D.~Janc and M.~Rosina,
%``The $T_{cc} = DD^*$ molecular state,''
Few Body Syst. \textbf{35}, 175-196 (2004)
doi:10.1007/s00601-004-0068-9
[arXiv:hep-ph/0405208 [hep-ph]].
%136 citations counted in INSPIRE as of 16 Aug 2022

%\cite{Li:2012ss}
\bibitem{Li:2012ss}
N.~Li, Z.~F.~Sun, X.~Liu and S.~L.~Zhu,
%``Coupled-channel analysis of the possible $D^{(*)}D^{(*)}, \overline{B}^{(*)}\overline{B}^{(*)}$ and $D^{(*)}\overline{B}^{(*)}$ molecular states,''
Phys. Rev. D \textbf{88}, no.11, 114008 (2013)
doi:10.1103/PhysRevD.88.114008
[arXiv:1211.5007 [hep-ph]].
%70 citations counted in INSPIRE as of 16 Aug 2022

%\cite{Kim:2018nqf}
\bibitem{Kim:2018nqf}
J.~Y.~Kim and H.~C.~Kim,
%``Electromagnetic form factors of singly heavy baryons in the self-consistent SU(3) chiral quark-soliton model,''
Phys. Rev. D \textbf{97}, no.11, 114009 (2018)
doi:10.1103/PhysRevD.97.114009
[arXiv:1803.04069 [hep-ph]].
%32 citations counted in INSPIRE as of 19 Nov 2022

%\cite{Kim:2020nug}
\bibitem{Kim:2020nug}
J.~Y.~Kim, H.~C.~Kim, M.~V.~Polyakov and H.~D.~Son,
%``Strong force fields and stabilities of the nucleon and singly heavy baryon $\Sigma_c$,''
Phys. Rev. D \textbf{103}, no.1, 014015 (2021)
doi:10.1103/PhysRevD.103.014015
[arXiv:2008.06652 [hep-ph]].
%14 citations counted in INSPIRE as of 19 Nov 2022

%\cite{Won:2022cyy}
\bibitem{Won:2022cyy}
H.~Y.~Won, J.~Y.~Kim and H.~C.~Kim,
%``Gravitational form factors of the baryon octet with flavor SU(3) symmetry breaking,''
[arXiv:2210.03320 [hep-ph]].
%1 citations counted in INSPIRE as of 19 Nov 2022
\end{thebibliography}
\end{document}